\documentclass{webofc}
\pdfoutput=1

\usepackage[varg]{txfonts}   
\usepackage[colorlinks=true
,urlcolor=magenta
,anchorcolor=green
,citecolor=red
,filecolor=green
,linkcolor=blue
,menucolor=green
,linktocpage=true
,pdfauthor={Daniel Grumiller, Raphaela Wutte}
,pdfa=true
]{hyperref}

\newcommand{\eq}[2]{\begin{equation} #1 \label{#2} \end{equation}}
\newcommand{\vp}{\varphi}
\DeclareMathOperator{\extdm}{d}
\newcommand{\extd}{\extdm \!}

\newcommand{\cJ}{{\cal J}}
\newcommand{\cK}{{\cal K}}

\begin{document}
\title{New entropy formula for Kerr black holes}

\author{\firstname{Hern\'an~A.} \lastname{Gonz\'alez}\inst{1}\fnsep\thanks{\email{hgonzale@hep.itp.tuwien.ac.at}} \and
        \firstname{Daniel} \lastname{Grumiller}\inst{1}\fnsep\thanks{\email{grumil@hep.itp.tuwien.ac.at}} \and
        \firstname{Wout} \lastname{Merbis}\inst{1}\fnsep\thanks{\email{merbis@hep.itp.tuwien.ac.at}}\and
        \firstname{Raphaela} \lastname{Wutte}\inst{1}\fnsep\thanks{\email{rwutte@hep.itp.tuwien.ac.at}}
}

\institute{Institute for Theoretical Physics, TU Wien, Wiedner Hauptstr.~8-10/136, A-1040 Vienna, Austria}

\abstract{%
We introduce a new entropy formula for Kerr black holes inspired by recent results for 3-dimensional black holes and cosmologies with soft Heisenberg hair. We show that also Kerr--Taub--NUT black holes obey the same formula.
}
\maketitle

\section{Introduction}\label{se:1}

Recently a new horizon entropy formula has emerged \cite{Afshar:2016wfy} 
\eq{
S=2\pi\,\big(J_0^+ + J_0^-\big)
}{eq:k1}
that is more universal than the Bekenstein--Hawking \cite{Bekenstein:1973ur, Hawking:1975sw} or Wald's \cite{Wald:1993nt, Iyer:1994ys} entropy formulas, albeit only in three dimensions. The quantities $J_0^\pm$ are zero-mode charges of $\mathfrak{u}(1)$ current algebras that appear to arise generically in near horizon descriptions.

Formula \eqref{eq:k1} was first derived for black holes in 3-dimensional Anti-de~Sitter space within Einstein gravity \cite{Afshar:2016wfy} inspired by related earlier discussions of near horizon boundary conditions \cite{Donnay:2015abr, Afshar:2015wjm} and the concept of soft hair \cite{Hawking:2016msc}. During the past 18 months the entropy formula \eqref{eq:k1} was shown to apply also to flat space cosmologies in Einstein gravity \cite{Afshar:2016kjj}, to black holes (flat space cosmologies) in higher spin theories \cite{Grumiller:2016kcp} (\cite{Ammon:2017vwt}) where Bekenstein--Hawking fails, and to higher derivative theories with gravitational Chern--Simons term \cite{Setare:2016vhy} where Wald's formula fails.

Besides the entropy itself also Cardyology simplifies \cite{Afshar:2016wfy, Perez:2016vqo, Afshar:2016kjj} including log-corrections \cite{Grumiller:2017jft}, while semi-classical considerations allow for Hardyology \cite{Afshar:2016uax, Sheikh-Jabbari:2016npa, Afshar:2017okz, Hajian:2017mrf}, by which we mean a counting of explicitly constructed semi-classical black hole microstates that combinatorially reduces to partitions of large integers. Nearly all results so far apply only to three spacetime dimensions, with the exception of \cite{Hajian:2017mrf} that applies Hardyology to extremal Kerr black holes. The phenomenologically most interesting non-extremal Kerr black hole \cite{Kerr:1963ud} was not discussed in detail so far.

The main purpose of this proceedings contribution is to make modest progress towards lifting the exciting results in three dimensions to four dimensions by providing an analog of the entropy formula \eqref{eq:k1} for non-extremal Kerr black holes. Our main result is a new entropy formula
\eq{
S_{\textrm{\tiny Kerr}} = 4\pi\,J_0^+ J_0^-
}{eq:k2}
that is equivalent to the Bekenstein--Hawking law or Wald's entropy, but expressed in terms of zero mode charges of $\mathfrak{u}(1)$ current algebras that are expected to appear in a suitable near horizon description of non-extremal Kerr.

This work is organized as follows.
In section \ref{se:2} we summarize briefly salient aspects of the black hole entropy \eqref{eq:k1} in three dimensions.
In section \ref{se:3} we present our new results for the entropy of non-extremal Kerr black holes.
In section \ref{se:new} we elaborate on inner horizon entropy, limits of Kerr and the generalization to Kerr--Taub--NUT.
In section \ref{se:4} we provide a brief outlook.

\section{Black hole entropy in three dimensions}\label{se:2}

In this section we review material that appeared in \cite{Afshar:2016wfy, Afshar:2016kjj}, whose conventions and notations we use. We start with the near horizon expansion of non-extremal black holes (or cosmologies).

The vicinity of non-extremal horizons allows an expansion around Rindler space
\eq{
\extd s^2=-\alpha^2r^2\,\extd t^2+\extd r^2+\gamma^2\,\extd\varphi^2+\cdots\
}{eq:k3}
where the ellipsis denotes terms that vanish as the radial coordinate tends to zero, $r\to 0$. The angular coordinate is periodic, $\varphi\sim\varphi+2\pi$. The exact solutions of Einstein's equations with negative cosmological constant $\Lambda=-1/\ell^2$ that asymptote to the near horizon metric \eqref{eq:k3} are given by 
\eq{
\extd s^2= -\alpha^2\ell^2\sinh^2(r/\ell)\,\extd t^2 + \extd r^2 +2\alpha\omega\ell^2\sinh^2(r/\ell)\,\extd t\extd\varphi  +\big(\gamma^2\cosh^2(r/\ell)-\omega^2\ell^2\sinh^2(r/\ell)\big)\,\extd\varphi^2 \,.
}{eq:k4}
As usual, for defining the theory it is crucial to declare which variations of the metric are allowed, i.e., to impose some consistent set of boundary conditions.

The near horizon boundary conditions of \cite{Afshar:2016wfy, Afshar:2016kjj} allow for arbitrary variations of the horizon radius function $\gamma$ and the rotation function $\omega$ but no variations of Rindler acceleration $\alpha$, i.e., $\delta\gamma\neq 0 \neq\delta\omega$ and $\delta \alpha=0$. For constant Rindler acceleration the equations of motion imply conservation of these functions in time, $\partial_t\gamma=0=\partial_t\omega$, which are near horizon analogs of the holographic Ward identities $\partial_\mp T^{\pm\pm}=0$ for the usual Brown--Henneaux boundary conditions \cite{Brown:1986nw}.

The canonical boundary charges associated with these choices are given by the Fourier modes
\eq{
J^\pm_n = \frac{1}{16\pi G}\,\oint \extd\vp\,e^{\pm in\vp}\,\big(\gamma \pm \omega\ell\big)
}{eq:k5}
where $G$ is the 3-dimensional Newton constant. Their algebra consists of two $\mathfrak{u}(1)$ current algebras
\eq{
[J_n^\pm,\,J_m^\pm] = \frac{\ell}{8G}\,n\,\delta_{n+m,\,0}\qquad [J^+_n,\,J^-_m] = 0\,.
}{eq:k6}
The zero mode charges $J_0^\pm$ commute with all generators $J_n^\pm$, which encodes algebraically the ``soft hair'' property. Since the algebra \eqref{eq:k6} can be mapped to infinite copies of the Heisenberg algebra frequently the name ``soft Heisenberg hair'' is used to distinguish from other soft hair constructions.

The Bekenstein--Hawking entropy is given by
\eq{
S_{\textrm{\tiny BH}} = \frac{2\pi\,\oint\gamma}{4G} = 2\pi\,\big(J_0^+ + J_0^-\big)
}{eq:k7}
which coincides with the result announced in \eqref{eq:k1}. While in Einstein gravity the result \eqref{eq:k7} is perhaps not too surprising, its simplicity is still remarkable and should be contrasted with the usual form of the entropy expressed in terms of Virasoro zero modes $L_0^\pm$ \cite{Cardy:1986ie, Strominger:1997eq, Carlip:1998qw},
\eq{
S_{\textrm{\tiny Cardy}} = 2\pi\,\Bigg(\sqrt{\frac{c L_0^+}{6}}+\sqrt{\frac{c L_0^-}{6}}\Bigg)
}{eq:k8}
with the Brown--Henneaux central charge
\eq{
c=\frac{3\ell}{2G} \,.
}{eq:angelinajolie}

More suprisingly, formula \eqref{eq:k1} applies also to flat space, where the usual Cardy-like formula is given by \cite{Bagchi:2012xr, Barnich:2012xq} ($L_0, M_0$ are BMS$_3$ zero modes and $c_M$ is the Barnich--Comp\`ere central charge \cite{Barnich:2006av})
\eq{
S_{\textrm{\tiny BMS}_3} = 2\pi\,L_0\,\sqrt{\frac{c_M}{2M_0}}
}{eq:k9}
and to situations where Bekenstein--Hawking fails, such as higher spin \cite{Henneaux:2010xg, Campoleoni:2010zq} black holes \cite{Gutperle:2011kf, Ammon:2012wc}, where even for the simplest case of spin-3 black holes the entropy looks fairly complicated \cite{Bunster:2014mua} ($L_0^\pm$ and $c$ are the same as in the spin-2 case; $W_0^\pm$ denotes the spin-3 zero mode charges with suitable normalization; for $W_0^\pm=0$ the Cardy formula \eqref{eq:k8} is recovered)
\eq{
S_{\textrm{\tiny spin\,3}}=2\pi\,\Bigg(\sqrt{\frac{cL_0^+}{6}}\cos\bigg(\frac13\,\arcsin\Big(W_0^+\big(L_0^+\big)^{-3/2}\Big)\bigg) + \sqrt{\frac{cL_0^-}{6}}\cos\bigg(\frac13\,\arcsin\Big(W_0^-\big(L_0^-\big)^{-3/2}\Big)\bigg)\Bigg)\,.
}{eq:k10}
Similar remarks and results apply to higher spin flat space \cite{Grumiller:2013mxa, Gonzalez:2013oaa} cosmologies \cite{Gary:2014mca, Matulich:2014hea, Riegler:2014bia}.

Let us note for later purposes that in the flat space case the algebra \eqref{eq:k6} appears naturally in an off-diagonal basis, $J_n = J_n^+ + J_{-n}^-$ and $K_n = \frac1\ell\,(J_n^+ - J_{-n}^-)$, such that
\eq{
[J_n,\,K_m] = \frac{1}{4G}\, n\,\delta_{n+m,\,0}
}{eq:k31}
and the entropy \eqref{eq:k1} reads
\eq{
S = 2\pi\, J_0\,.
}{eq:k32}

The algebraic reasons behind these surprising results are Sugawara-type of relations between the near horizon generators $J_n^\pm$ and the usual asymptotic variables, like Virasoro generators, BMS$_3$ generators or higher spin generators. For example, the Virasoro generators $L_n^\pm$ emerge uniquely from a Miura-type of transformation (spelled out in \cite{Afshar:2016wfy}) that leads to a Sugawara construction
\eq{
L_n^\pm = \frac{4G}{\ell}\,\sum_p J_{n-p}^\pm J_p^\pm + in J_n^\pm
}{eq:k11}
with a precise twist-term that allows to recover the Brown--Henneaux central charge \eqref{eq:angelinajolie}. Inserting this central charge together with the vacuum expectation value of the zero-mode generator
\eq{
L_0^\pm = \frac{4G}{\ell}\, \big(J_0^\pm\big)^2
}{eq:k12}
into the Cardy formula \eqref{eq:k8} recovers formula \eqref{eq:k1}. In the spin-3 case the Sugawara-type construction involves tri-linears in the spin-2 and spin-3 generators $J_{(2,3)\,n}^\pm$ \cite{Grumiller:2016kcp} which in a similar way allows to recover formula \eqref{eq:k1} from the complicated expression \eqref{eq:k10} and vice versa. A summary of some of these algebraic constructions is provided in \cite{Afshar:2016kjj}.

Besides these unexpected simplifications of the results for entropy, at least in three dimensions one can go further and exploit the near horizon theory for a microstate counting (including log corrections), see \cite{Grumiller:2017jft} and references therein. Moreover, based on the $\mathfrak{u}(1)$ current algebras \eqref{eq:k6} a specific proposal for all microstates of semi-classical BTZ black holes was provided in \cite{Afshar:2016uax, Afshar:2017okz}.

While it is not yet clear how these more advanced considerations generalize to higher dimensions, the more basic ones reviewed here show at least that there is a simple and fairly universal entropy law \eqref{eq:k1} in three dimensions that often considerably simplifies the results and paves the way for a deeper analysis of black hole entropy. We show in the next section that something similar can be achieved for non-extremal Kerr black holes.

\section{Kerr black hole entropy}\label{se:3}

\subsection{Bekenstein--Hawking entropy}\label{se:3.1}

Kerr black holes are characterized uniquely by their mass $M$ and angular momentum $J = aM$, in terms of which their Bekenstein--Hawking entropy reads (we set Newton's constant to unity)
\eq{
S_{\textrm{\tiny Kerr}} = 2\pi\,M^2\,\Big(1 + \sqrt{1 - a^2/M^2}\Big)\,.
}{eq:k13}

\subsection{Near horizon symmetry algebra by Donnay et al.}\label{se:3.2}

To continue with our agenda we need some analog of the near horizon boundary conditions reviewed in section \ref{se:2}, which so far do not exist. Thus, we use instead the near horizon boundary conditions introduced in \cite{Donnay:2015abr} (see also \cite{Donnay:2016ejv}), which lead to a symmetry algebra whose non-vanishing commutators are displayed below.
\begin{subequations}
\label{eq:k14}
\begin{align}
 [L_n^\pm,\,L_m^\pm] &= (n-m)\,L_{n+m}^\pm \\
 [L_n^+,\,T_{(m,\,k)}] &= -m\,T_{(n+m,\,k)} \\
 [L_n^-,\,T_{(m,\,k)}] &= -k\,T_{(m,\,n+k)} 
\end{align}
\end{subequations}
The Witt generators $L_n^\pm$ produce ``superrotations'' and the remaining generators $T_{(n,\,m)}$ ``supertranslations''. Note, however, that the algebra \eqref{eq:k14} differs from the Barnich--Troessaert generalization \cite{Barnich:2009se, Barnich:2010eb} of BMS \cite{Bondi:1962, Sachs:1962}. 

\subsection{Sugawara deconstruction}\label{se:3.3}

There are three key observations for our purposes. The first one, made in \cite{Donnay:2015abr}, relates the supertranslation double-zero-mode to the Kerr black hole entropy \eqref{eq:k13},\footnote{\label{fn:1} For convenience our normalization of $T_{(0,\,0)}$ differs from the one in \cite{Donnay:2015abr} by a factor of surface gravity and a factor $2$.}
\eq{
S_{\textrm{\tiny Kerr}} = 4\pi\,T_{(0,0)}
}{eq:k15}
The second one (made in the same paper) is that for the Kerr solution the Witt zero mode generators, up to an imaginary factor, are given by the Kerr angular momentum.
\eq{
L_0^\pm = \pm \frac{i}{2}\,J
}{eq:k18}
The third one, made in \cite{Afshar:2016uax}, provides a ``Sugawara-deconstruction'' of the algebra \eqref{eq:k14} in terms of four $\mathfrak{u}(1)$ current algebras (commutators not displayed vanish)
\eq{
[\cJ_n^\pm,\,\cJ_m^\pm] = - [\cK_n^\pm,\,\cK_m^\pm] = \frac{n}{2}\, \delta_{n+m,\,0} 
}{eq:k16}
namely
\eq{
T_{(n,\,m)} = \big(\cJ_n^+ + \cK_n^+\big)\big(\cJ_m^- + \cK_m^-\big) \qquad\qquad L_n^\pm = \sum_p \big(\cJ_{n-p}^\pm + \cK_{n-p}^\pm\big) \big(\cJ_p^\pm - \cK_p^\pm\big)\,.
}{eq:k17}

The ``deconstruction'' above is of course not unique and we can change to a more convenient basis, which we do now. Defining the generators
\eq{
J_n^\pm := \cJ^\pm_n + \cK^\pm_n \qquad\qquad  K_n^\pm := \cJ^\pm_n - \cK^\pm_n 
}{eq:k25}
yields an algebra consisting of two copies of the 3-dimensional flat space near horizon symmetry algebra \eqref{eq:k31} (again all commutators not displayed vanish)
\eq{
[J_n^\pm,\,K_m^\pm] = n\, \delta_{n+m,\,0} \,.
}{eq:k26}
The generators \eqref{eq:k17} are given by the bilinears
\eq{
T_{(n,\,m)} = J_n^+ J_m^- \qquad\qquad L_n^\pm = \sum_p J_{n-p}^\pm K_p^\pm \,.
}{eq:k30}

\subsection{Determination of zero modes}\label{se:inbetween}

Comparing with \cite{Donnay:2015abr} for the non-extremal Kerr solution the four zero modes of the new generators have to obey three algebraic constraints (determined by their relations to $L_0^\pm$ and $T_{(0,\,0)}$). To solve uniquely for these zero modes we need a fourth condition. We impose
\eq{
J_0^+ + J_0^- + K_0^+ + K_0^- \stackrel{!}{=} 2M 
}{eq:k33}
based on the rationale that this chirally symmetric sum of zero modes should not see the angular momentum, while dimensional analysis shows it must be linear in mass. The least obvious aspect of the postulate \eqref{eq:k33} is the factor $2$ on the right hand side. This factor is fixed uniquely by demanding that in the Schwarzschild limit of vanishing angular momentum, $J\to 0$, the zero modes agree with each other pairwise, $J_0^+ = J_0^-$ (the other relation, $K_0^+=K_0^-$, holds automatically in this limit since both quantities tend to zero). Solving the algebraic equations \eqref{eq:k33}, $T_{(0,\,0)}=J_0^+ J_0^-$ and $L_0^\pm = J_0^\pm K_0^\pm$ establishes uniquely results for the zero modes of the new generators.
\eq{
J_0^\pm = \frac{1}{2}\,\big(M + \sqrt{M^2-a^2} \pm i a\big) \qquad\qquad
K_0^\pm = \frac{1}{2}\,\big(M - \sqrt{M^2-a^2} \pm i a\big) 
}{eq:k27}

\subsection{Results for non-extremal Kerr in terms of \texorpdfstring{$\boldsymbol{\mathfrak{u}(1)}$}{u(1)} zero modes}\label{se:3.4}

We express now standard Kerr parameters in terms of the zero mode charges \eqref{eq:k27}. The black hole and Cauchy horizon radii 
\eq{
r_+ = J_0^+ + J_0^- \qquad\qquad r_- = K_0^+ + K_0^- 
}{eq:k22}
as well as the black hole mass 
\eq{
M = J_0^+ + K_0^- = J_0^- + K_0^+
}{eq:k28}
are linear in the zero mode charges, while angular momentum
\eq{
J = -2i\, J_0^+ K_0^+ = 2i\, J_0^- K_0^- 
}{eq:k23}
and, equivalently, the Virasoro zero modes \eqref{eq:k18}
\eq{
L_0^\pm = J_0^\pm K_0^\pm 
}{eq:k29}
are bilinear in them.

The Bekenstein--Hawking entropy \eqref{eq:k15} is also bilinear in the zero modes.
\eq{
\boxed{
S_{\textrm{\tiny Kerr}} = 4\pi\,J_0^+ J_0^-
}
}{eq:k20}
The new entropy formula \eqref{eq:k20} is the main result of this work. The non-trivial aspect is not just that the entropy can be written as in \eqref{eq:k20}, but rather that the quantities therein, $J_0^\pm$, are (linear combinations of) zero modes of $\mathfrak{u}(1)$ generators that are related in a Sugawara-like way to the near horizon charges \eqref{eq:k30} discovered in \cite{Donnay:2015abr}. As compared to the 3-dimensional flat space result \eqref{eq:k32} the entropy now contains a product of two zero-mode charges instead of a single one. The factor $4\pi$ in \eqref{eq:k20} is naturally interpreted as volume of the round unit 2-sphere.

\section{Elaborations}\label{se:new}

In this section we discuss some elaborations. We address inner horizon entropy in section \ref{se:3.5}, mention Schwarzschild and extremal limits in section \ref{se:3.6}, and present the generalization to Kerr--Taub--NUT in section \ref{se:tn}. We comment on some further interesting generalizations in the concluding section \ref{se:4}.

\subsection{Inner horizon entropy}\label{se:3.5}

The ``inner horizon entropy'' defined and used in \cite{Larsen:1997ge, Cvetic:1997uw, Cvetic:2010mn, Castro:2012av, Detournay:2012ug} is bilinear in the other pair of zero modes [the ones not appearing in the new entropy formula \eqref{eq:k20}].
\eq{
S_{\textrm{\tiny inner}} = 2\pi\,K_0^+ K_0^- = 2\pi\,M^2\,\Big(1-\sqrt{1-a^2/M^2}\Big)
}{eq:k21}
The result above suggests that the generators $J_n^\pm$ are associated with the outer (black hole) horizon, while the generators $K_n^\pm$ are associated with the inner (Cauchy) horizon. 
(Note that one can define generators $\tilde{T}_{(n,m)} = K_n^+ K^-_m$ that also satisfy \eqref{eq:k14}, but with $T$ replaced by $\tilde{T}$, which can be interpreted as generating supertranslations on the inner horizon.) 
This suggestion is confirmed by the expressions for outer and inner horizon radii \eqref{eq:k22}. As may have been anticipated, angular momentum \eqref{eq:k23} is a quantity that involves information about both horizons. The same is true for the mass \eqref{eq:k28}.

\subsection{Schwarzschild and extremal Kerr}\label{se:3.6}

For Schwarzschild the inner horizon charges vanish, $K_0^\pm=0$. Other than that all the results of section \ref{se:3} apply. 

By contrast, in the extremal limit the outer and inner horizon charges are identical, $J_0^\pm = K_0^\pm$, which physically makes sense as both horizons then coalesce to a single one. In that case the new entropy formula \eqref{eq:k20} by virtue of the angular momentum formulas \eqref{eq:k23} can be rewritten as
\eq{
S_{\textrm{\tiny EK}} = 4\pi\,\sqrt{J_0^+ K_0^+ J_0^- K_0^-} = 2\pi\,|J| 
}{eq:k34} 
which coincides with the near horizon extremal Kerr result \cite{Guica:2008mu, Compere:2012jk}. Note, however, that the boundary conditions of \cite{Donnay:2015abr} on which our construction is based solely apply to non-extremal Kerr black holes, so one has to be careful with this limit. See \cite{Hajian:2017mrf} for more on extremal Kerr entropy in a near horizon context, including a proposal for its microstates.

\subsection{Kerr--Taub--NUT}\label{se:tn}

We consider now a generalization to Kerr--Taub--NUT \cite{Taub:1950ez, Newman:1963yy} whose metric in Boyer--Lindquist coordinates is given by
\eq{
\extd s^2 = - \frac{\Delta}{\Sigma} \left(\extd t - (a \sin^2 \theta - 2n \cos \theta)\extd\varphi \right)^2 + \frac{\Sigma}{\Delta}\,\extd r^2 + \Sigma\,\extd\theta^2 + \frac{\sin^2\theta}{\Sigma} \left( a \extd t^2 -(r^2+ a^2 + n^2)\extd\varphi \right)^2
}{eq:KTN}
with Kerr parameter $a=J/M$, nut charge $n$ and the two functions
\eq{
\Delta = r^2 - 2 Mr +a^2 - n^2 \qquad\text{and} \qquad 
\Sigma = r^2 + \left( n + a \cos \theta \right)^2\,.
}{eq:SDdef}
Rather than repeating in detail the analysis we did for Kerr we merely state the result for the zero mode charges.
\eq{
J_0^\pm = \frac{1}{2}\,\Big(M + \sqrt{M^2-a^2 + n^2} \pm i a\Big) \qquad K_0^\pm =\frac12\,\Big( M - \sqrt{M^2-a^2 + n^2} \pm i a\Big)
}{eq:k35}
Besides mass $M$ and angular momentum $J$ the zero mode charges \eqref{eq:k35} depend also on the nut charge $n$.

Repeating the analysis of section \ref{se:3.4} yields expressions for the horizon radii
\eq{
r_+ = J_0^+ + J_0^- = M + \sqrt{M^2-a^2+n^2} \qquad r_- = K_0^+ + K_0^- = M - \sqrt{M^2-a^2+n^2} 
}{eq:k36}
and the mass.
\eq{
M = J_0^+ + K_0^- = J_0^- + K_0^+ 
}{eq:k39}
The main change as compared to Kerr is that the left and right Witt zero mode charges $L_0^\pm=J_0^\pm K_0^\pm$ do not add up to zero, $L_0^+ + L_0^-\neq 0$, but instead yield the square of the nut charge.
\eq{
-\frac{n^2}{2}  = J_0^+ K_0^+ + J_0^- K_0^-
}{eq:k41}
The angular momentum $J = a M$ is determined from the difference of the Witt zero mode charges. 
\eq{
iJ  = J_0^+ K_0^+ - J_0^- K_0^-
}{eq:k40}
The Bekenstein--Hawking entropy is again compatible with our main result \eqref{eq:k20}.
\eq{
S_{\textrm{\tiny Kerr-Taub-NUT}}=4\pi J_0^+ J_0^- = 2\pi\,\Big(M^2 +n^2 + M \sqrt{M^2-a^2+n^2}\Big)
}{eq:k37}
For completeness we also state the corresponding result for inner horizon entropy.
\eq{
S_{\textrm{\tiny inner}}=4\pi K_0^+ K_0^- = 2\pi\,\Big(M^2 +n^2 - M \sqrt{M^2-a^2+n^2}\Big)
}{eq:k38}

\subsection{Summary of relations between \texorpdfstring{$\boldsymbol{\mathfrak{u}(1)}$}{u(1)} zero mode charges and black hole parameters}\label{se:table}

Table \ref{tab:1} summarizes various linear and bilinear relations between the $\mathfrak{u}(1)$ zero mode charges $J_0^\pm$, $K_0^\pm$ and black hole quantities.
\renewcommand{\arraystretch}{1.35}
\begin{table}[h!]
\label{tab:1}
\begin{center}
\begin{tabular}{|l|l|l|}\hline
zero mode combination & black hole quantity & physical interpretation \\ \hline
$J_0^+ + J_0^-$ & $r_+$ & black hole radius \\
$K_0^+ + K_0^-$ & $r_-$ & Cauchy horizon radius \\
$J_0^+ + K_0^+$ & $M$ & black hole mass \\
$J_0^+ - K_0^+$ & $\sqrt{M^2-a^2+n^2}$ & extremality parameter \\
$J_0^+ - J_0^-$ & $ia=iJ/M$ & Kerr parameter\\ \hline
$\boldsymbol{J_0^+ J_0^-}$ & $\boldsymbol{S/(4\pi)}$ & {\bf black hole entropy} \\ \hline
$K_0^+ K_0^-$ & $S_{\textrm{\tiny inner}}/(4\pi)$ & ``inner horizon entropy'' \\
$J_0^+ K_0^+ - J_0^- K_0^-$ & $iJ$ & angular momentum \\
$J_0^+ K_0^+ + J_0^- K_0^-$ & $-n^2/2$ & nut charge (squared)\\
\hline
\end{tabular}
\end{center}
\caption{List of linear and bilinear relations between $\mathfrak{u}(1)$ zero mode charges $J_0^\pm$, $K_0^\pm$ and black hole quantities}
\end{table}

\section{Outlook}\label{se:4}

Our main result \eqref{eq:k20} is a reformulation of the usual Kerr black hole entropy. As indicated, we view this as the first modest step in a more ambitious program, namely to provide near horizon boundary conditions, discuss associated symplectic and/or asymptotic symmetries, derive Cardyology and propose Hardyology for non-extremal Kerr. The endgame of this research avenue is to provide an explicit set of all microstates for semi-classical non-extremal Kerr black holes and to discuss consequences for black hole evaporation, (avoidance of) information loss and related semi-classical black hole puzzles.

In four and five spacetime dimensions Cveti\v{c} and collaborators have found that the product of black hole and ``inner horizon'' entropies is some integer multiple of $4\pi^2$, see \cite{Cvetic:2010mn} and references therein as well as papers by Ansorg and collaborators, e.g.~\cite{Ansorg:2009yi}. Our results \eqref{eq:k20} and \eqref{eq:k21} actually suggest that already the individual entropies are some integer multiples of $2\pi$, provided the zero mode charges $J_0^\pm$ and $K_0^\pm$ are quantized in the integers. Results for three-dimensional black holes suggest that this is the case \cite{Afshar:2016uax, Afshar:2017okz}, but clearly it would be desirable to verify this in four dimensions.

Another interesting research direction is to relate our results to the hidden conformal symmetry for Kerr black holes, see \cite{Castro:2010fd} and references therein. While we do not see direct evidence for conformal symmetry, our $\mathfrak{u}(1)$ currents $J_n^\pm$ and $K_n^\pm$ can be Sugawara-combined to Virasoro generators, which then indirectly generate conformal symmetry, see \cite{Afshar:2017okz} for a summary of the relevant algebraic statements.

The new entropy formula for Kerr \eqref{eq:k20} and some of the directions alluded to above should be generalizable to other black holes in four dimensions with cosmological constant, electromagnetic sources and acceleration (e.g.~Carter--Kottler a.k.a.~Kerr-\mbox{(anti-)}de~Sitter \cite{Kottler:1918, Carter:1973}, Kerr--Newman \cite{Newman:1965my, Newman:1965tw}, accelerated black holes \cite{Appels:2016uha}, Pleba\'nski--Demia\'nski \cite{Plebanski:1976gy} or its cosmological generalization \cite{Griffiths:2005qp}, and black holes with toroidal or higher genus horizons \cite{Brill:1997mf}) as well as to black holes (and possibly other black objects such as black rings \cite{Emparan:2001wn}) in more than four spacetime dimensions. Studying these generalization would also allow to verify if the simple entropy formula \eqref{eq:k20} is again as universal as its 3-dimensional pendant \eqref{eq:k1}.

Besides the four- and higher-dimensional applications mentioned above there is still some interesting work to be done in three spacetime dimensions. Namely, so far formula \eqref{eq:k1} was verified solely for locally maximally symmetric metrics, such as Ba\~nados--Teitelboim--Zanelli black holes \cite{Banados:1992wn,Banados:1992gq} or flat space cosmologies \cite{Cornalba:2002fi, Cornalba:2003kd}. However, the near horizon approximation \eqref{eq:k3} should be valid for any horizon, even if the solution is not maximally symmetric. It would be interesting to verify whether our general conclusions still apply to such scenarios, for instance to warped black holes, see e.g.~\cite{Anninos:2008fx}. Further open issues in three spacetime dimensions are listed in the concluding section of \cite{Afshar:2016kjj}.

Whether all these research avenues will turn out to be fruitful is unclear. However, in our opinion it is exciting that there is now a new approach available --- based on near horizon considerations and soft Heisenberg hair --- to address macro- and microscopic aspects of black hole entropy.

\section*{Acknowledgments}

DG thanks Soo-Jong Rey and the organizers of the 13$^{\textrm{th}}$ International Conference on Gravitation, Astrophysics, and Cosmology and the 15$^{\textrm{th}}$ Korea-Italy Relativistic Astrophysics Symposium for the invitation and the participants for enjoyable discussions. We thank Hamid Afshar, Martin Ammon, Steph\'ane Detournay, Laura Donnay, Gaston Giribet, Alfredo Perez, Miguel Pino, Stefan Prohazka, Max Riegler, Shahin Sheikh-Jabbari, David Tempo, Ricardo Troncoso and Hossein Yavartanoo for collaboration on various aspects of near horizon symmetries, soft Heisenberg hair, black hole entropy and semi-classical black hole microstates. Additionally, we thank Glenn Barnich, Steve Carlip, Geoffrey Comp\`ere, Jakob Salzer, Friedrich Sch\"oller and Andy Strominger for discussions.

This work was supported by the Austrian Science Fund (FWF), projects P~27182-N27 and P~28751-N27. 

\addcontentsline{toc}{section}{References}

\end{document}